\documentclass[fleqn,usenatbib]{mnras}
\usepackage[T1]{fontenc}

\DeclareRobustCommand{\VAN}[3]{#2}
\let\VANthebibliography\thebibliography
\def\thebibliography{\DeclareRobustCommand{\VAN}[3]{##3}\VANthebibliography}

\usepackage{graphicx}	% Including figure files
\usepackage{amsmath}	% Advanced maths commands
\usepackage{amssymb}
\usepackage{ulem}
\usepackage{tikz}
\usepackage{multirow}
\usepackage{newtxtext,newtxmath}

\title[LFQPOs in GX\,339--4]{Evolution of low frequency quasi-periodic oscillations in GX\,339--4 during its 2021 outburst using {\it AstroSat} data}

\author[Mondal et al.]{
Santanu Mondal$^{1}$\thanks{E-mail:santanu.mondal@iiap.res.in (SM)}, Anirudh Salgundi$^{2,1}$\thanks{E-mail:salgundi.anirudh@gmail.com (AS)}, Debjit Chatterjee$^{1}$, Arghajit Jana$^{3}$\thanks{E-mail:argha0004@gmail.com(AJ)}, H.-K. Chang$^{3}$, Sachindra Naik$^{4}$
\\
$^{1}$Indian Institute of Astrophysics, II Block Koramangala, Bangalore 560034, India \\
$^{2}$Indian Institute of Technology Bombay, Powai 400076, India \\
$^{3}$Institute of Astronomy, National Tsing Hua University, Hsinchu 300044, Taiwan\\
$^{4}$Astronomy and Astrophysics Division, Physical Research Laboratory, Navrangpura, Ahmedabad - 380009, Gujarat, India
}

\date{Accepted XXX. Received YYY; in original form ZZZ}

\begin{document}
\label{firstpage}
\pagerange{\pageref{firstpage}--\pageref{lastpage}}
\maketitle

\begin{abstract}
The black hole X-ray binary GX\,339-4 showed an X-ray outburst during 2021. The {\it AstroSat} captured this outburst when the source entered into the intermediate flux state while the count rate was declining. The source showed an alternating flux profile in a timescale of $\lesssim$100 ks, where the hard energy band was more variable than the soft band. The energy-dependent timing study showed that the observed quasi-periodic oscillation (QPO) was prominent in the low energy bands, with its nearly sub-harmonic and harmonic components. These components appear and disappear with time, as observed in the orbit-wise QPO study. The Q-value, fractional rms, and 4.8-5.6 Hz frequency infer the QPOs as type-B and the spectral state as soft intermediate. The rms spectra of all orbits exhibiting QPOs show an increase in amplitude till $\sim$ 10 keV, beyond which it starts decreasing. This may indicate that $\sim$ 10 keV photons contributed relatively more in QPOs than other energy band photons. The Lorentzian normalization of the type-B QPO in different energy bands is consistent with the 10 keV peak. The energy-dependent time lag is complex and could be associated with the Comptonizing corona or jet. Finally, we discuss possible reasons behind the origin of different timing properties observed. 
\end{abstract}

% Don't make up new oneThe part that I found more noteworthy and novel is the usage of the JETCAF model on a ULX spectrum.s.

%%%%%%%%%%%%%%%%%%%%%%%%%%%%%%%%%%%%%%%%%%%%%%%%%%
%%%%%%%%%%%%%%%%% BODY OF PAPER %%%%%%%%%%%%%%%%%%
\section{introduction}
A transient black hole X-ray binary (BHXB) spends most of its time in quiescence and occasionally shows X-ray outbursts. During the outburst, the BHXBs show rapid variabilities in both spectral and temporal properties. The X-ray spectrum of a BHXB can be modeled by a multi-color disc blackbody (MCD) and a power-law tail (PL). The MCD is believed to arise from a geometrically thin and optically thick accretion disc \citep{ShakuraSunyaev1973A&A....24..337S}, while the PL tail is believed to originate from a hot electron cloud known as Compton cloud or corona \citep[e.g.,][]{HM93,ChakTitarchuk1995ApJ...455..623C,Done2007}. The soft thermal photons emerging from the disc undergo inverse-Compton scattering in the Compton cloud, producing a hard Comptonized power-law tail \citep{ST80,ST85}.

The fast variability is observed in the power density spectra (PDS) of the light curve. The band-limited noise with a flat profile in $\nu-P_{\nu}$ space with a break frequency ($\nu_{\rm b}$) characterize the PDS \citep[e.g.,][]{vanderklis1989,vanderklis1994}. The noise profile of the PDS can be approximated by one or more Lorentzian functions \citep[e.g.,][]{Nowak2001,BelloniEtal2005A&A...440..207B}. Often, a peaked-noise or quasi-periodic oscillation (QPO) is seen in the PDS of the BHXBs. The low-frequency QPO (LFQPO; $\nu_{\rm QPO} = 0.1-30$~Hz) can be classified as type-A, type-B or type-C, depending on the Q-value (Q=$\nu/\Delta \nu$, $\nu$ and $\Delta \nu$ are the centroid frequency and full-width half maximum, FWHM), amplitude (\% rms), and $\nu$ \citep[see ][ and references therein]{CasellaEtal2005ApJ...629..403C}. 

The correlation between the spectral and timing properties are observed in the hardness-intensity diagram \citep[HID; e.g.,][]{HomanEtal2001ApJS..132..377H}, accretion rate ratio-intensity diagram \citep[ARRID][]{MondalEtal2014ApJ...786....4M,Janaetal2016}, rms-intensity diagram \cite[RID;][]{Munoz-Darias2011} or hardness ratio-rms diagram \cite[HRD;][]{BelloniEtal2005A&A...440..207B}. An outbursting BHXB exhibits different spectral states during the outburst, which is seen in the different branches of the HID, ARRID, RID, or HRD. Generally, the BHXB goes through low-hard state (LHS), hard-intermediate state (HIMS), soft-intermediate state (SIMS), and high soft state (HSS) during the outburst \citep[e.g.,][]{RemillardMcClin2006ARA&A..44...49R,NandiEtal2012A&A...542A..56N,AJ2022}. A detailed study on the spectral state evolution has been made by several groups for different outbursting BHXBs \citep[e.g.,][]{RemillardMcClin2006ARA&A..44...49R,ChatterjeeEtal2016ApJ...827...88C}.

The physical mechanism of LFQPO is not well understood. Several models have been proposed to explain the origin of the LFQPO in BHXBs, including magneto-acoustic waves \citep[e.g.,][]{Titarchuk1998, CabanacEtal2010MNRAS.404..738C}, spiral density waves \citep[e.g.,][]{varniere2002,varniere2012}, Lense-Thirring precession \citep[e.g.,][]{StellaEtal1999ApJ...524L..63S,IngramEtal2009MNRAS.397L.101I}, shock oscillation model \citep[e.g.,][]{MolteniEtal1996ApJ...457..805M,ChakrabartiEtal.POS2005,chakrabartiEtal2008A&A...489L..41C}. Most models do not consider the spectral properties of the BHXBs while explaining the LFQPOs. However, one needs a general comprehensive model to explain both spectral and timing properties physically. The shock oscillation model is one such model that can explain both spectral and timing properties in a single framework \citep{ChakrabartiEtal2015MNRAS.452.3451C,ChatterjeeEtal2016ApJ...827...88C}.

Transient BHXB GX~339--4 went into an X-ray outburst in January 2021 that lasted for about 10 months \citep{Corbel2021,Garcia2021}. {\it AstroSat} observed the source for a total of 600 ks between 30 March 2021 and 6 April 2021 when the source was in the intermediate state. We studied the timing and spectral properties of the source in detail using the data from the {\it AstroSat} observations. We present the timing properties of GX~339--4 during the intermediate state in this paper. The spectral properties will be presented elsewhere. The paper is organized in the following way. In \S2, the data reduction and analysis process are presented. In \S3, we present the results of the temporal analysis. Finally, in \S4, we discuss our results.

\section{Observations and data analysis}
{\it AstroSat} \citep{SinghEtal2014SPIE.9144E..1SS, AgrawalEtal2017JApA...38...30A} is India’s first multi-wavelength astronomical observatory, which contains five instruments onboard: Soft X-ray Telescope (SXT), Large Area X$-$ray Proportional Counter (LAXPC), Cadmium Zinc Telluride Imager (CZTI), a Scanning Sky Monitor (SSM) and an Ultra-Violet Imaging Telescope (UVIT). Among them, the LAXPC is the ideal instrument to probe rapid time variability studies of X-ray binaries with moderate spectral capabilities. The high sensitivity and medium resolution spectral capability of SXT in the $0.3-8$ keV energy band are useful for broadband spectral studies simultaneously with LAXPC. These capabilities provided a unique opportunity to investigate the spectro-timing properties of several X-ray binaries (including new transients), particularly BHXRBs using {\it AstroSat}.

LAXPC has three units such as LAXPC10, LAXPC20, and LAXPC30, operating in the $3-80$ keV energy band. The three units have a total effective area of $\sim 6000~\rm cm^{2}$ in 5$-$20 keV and record X-ray events with a time resolution of $10\,\rm \mu s$ \citep{Yad16, Yad16a, Antia2017, AgrawalEtal2017JApA...38...30A}. We used the Event Analysis (EA) mode data of the source and reduced the Level~1 data using the LAXPC software ({\tt LaxpcSoft}) to generate the Level~2 cleaned event files. Among the three units, LAXPC10 and LAXPC30 showed abnormal behaviors (gain change, low gain, and gas leakage). In addition, the LAXPC 30 has no longer been operational since 8 March 2018. Thus, for the present study, we have used the archival data of the \textit{AstroSat} observations of GX\,339--4 between 30 March and 6 April 2021 in the $3-80$ keV energy range from the LAXPC20 instrument. Further details of the observations are mentioned in \autoref{tab:ObsLog}. 

The light curves were extracted using the task {\fontfamily{qcr}\selectfont laxpc\_make\_lightcurve}\footnote{{\it AstroSat} Science Support Cell:  http://astrosat-ssc.iucaa.in/laxpcData}, and are background subtracted. The background estimation was done based on the blank sky observations, closest to the time of observation of the source \citep[see][for details]{Antia2017}. We divided the entire energy band (3$-$30 keV) combining all 74 orbits ($\sim$ 7 days) into several segments with narrow energy bands, namely 3$-$5.5 keV, 5.5$-$10 keV, 10$-$15 keV, 15$-$20 keV, and 20$-$30 keV and generated PDS using {\fontfamily{qcr}\selectfont laxpc\_rebin\_power} task for each of them. As the background is dominated above 25 keV, we limited our PDS analysis up to 20 keV. We applied fast-Fourier transform (FFT) to the 0.01 sec time binned light curve. The light curves were divided into some intervals, where each interval consisted of 8192 bins. Then, we produced Poisson noise subtracted PDS for each of these segmented light curves and were averaged to make a single PDS for each observation. A geometrical rebinning with step 0.02 is used for generating PDS. Fractional rms is used to normalize all PDS. We have fitted the 5.5-10 keV band QPOs (see the next section) using multiple Lorentzian components and a power-law component: one with a narrow and strong component, which fits the fundamental QPO ($\nu_{\rm qpo}^{\rm f}$), and the other two with relatively weaker components fitting the sub-harmonic ($\nu_{\rm qpo}^{\rm s}$), harmonic ($\nu_{\rm qpo}^{\rm h}$) QPOs, and the power-law component to model the broadband noise. The power-law normalization of $2.96\pm0.39\times 10^{-4}$ is measured.

Further, we look for the evolution of $\nu_{\rm qpo}$ during the {\it AstroSat} observation period. For that, we analyzed data from each orbit separately and searched for QPOs. %We used the same geometrical rebinning as above to generate the PDS. Here, we considered the $3-25$ keV energy band light curves. 
Out of 74 orbits, only 11 orbits showed the presence of QPOs with harmonic and sub-harmonic components. We fitted them using {\it Lorentzian} model and extracted the QPO parameters. The model fitted parameters for the 11 orbits are shown in \autoref{tab:qpo_var}. All uncertainties correspond to 90\% confidence level.  We generally considered QPO if a peaked noise was detected at three sigma level. Besides the first 11 orbits, no peaked noise was observed in the PDS.

We estimated the energy-dependent time lag using {\fontfamily{qcr}\selectfont laxpc\_find\_freqlag} for the combined 74 orbits. To generate the time-lag spectra, {\it LAXPC} subroutine provides inputs such as the frequency resolution ($\Delta f$) and frequency at which time-lag has to be computed ($f=5.06$ Hz in our case). The subroutine first generates PDS in different energy bands (between 3-32 keV with energy band size 2 keV), with one of the bands set as reference (9-11 keV as the rms amplitude is maximum in this band). It then estimates the phase of the cross-correlation function for each energy band with the reference band. The time-lag was calculated by dividing the phase-lag by $2\pi f$. The detail description on time-lag calculation \citep[][]{NowakEtal1999ApJ...517..355N} and {\it LAXPC} subroutine can be found in \citet[][and references therein]{Mis17,HussainEtal2023MNRAS.525.4515H}. The coherence factor varies between 0.1 to 3.5. To estimate the rms amplitude spectra, a 0.1-50 Hz frequency range is considered.

\begin{table}
    \centering
    \caption{Observation log of {\it AstroSat} data. In the second panel, only orbits (``O") that exhibit QPOs are mentioned.}
    \begin{tabular}{{|c|c|c|c|}}
    \hline
    \textbf{Obs. ID} & \textbf{Start time} & \textbf{Stop time} \\
                  & MJD & MJD \\
    \hline
    T03\_291T01\_9000004278  &  59303 &  59310 & \\
    \hline
O29750 &59303.044 & 59303.091\\ 
O29755 &59303.073 &59303.178 \\
O29756 &59303.181 &59303.516 \\
O29758 &59303.516 &59303.564 \\
O29759 &59303.589 &59303.637 \\
O29760 &59303.649 &59303.742 \\
O29761 &59303.732 &59303.815 \\
O29763 &59303.877 &59303.961 \\
O29764 &59303.950 &59304.033 \\
O29765 &59304.004 &59304.106 \\
O29770 &59304.095 &59304.318\\
\hline
    \end{tabular}
    \label{tab:ObsLog}
\end{table}

\section{Results and Discussions}
\begin{figure}
\hspace{-0.3cm}
\begin{tikzpicture}
\draw (0, 0) node[inner sep=0] {\raisebox{0.0cm}{\includegraphics[height=9truecm,width=9truecm,trim={0.0cm 2.0cm 2.5cm 3.0cm}, clip]{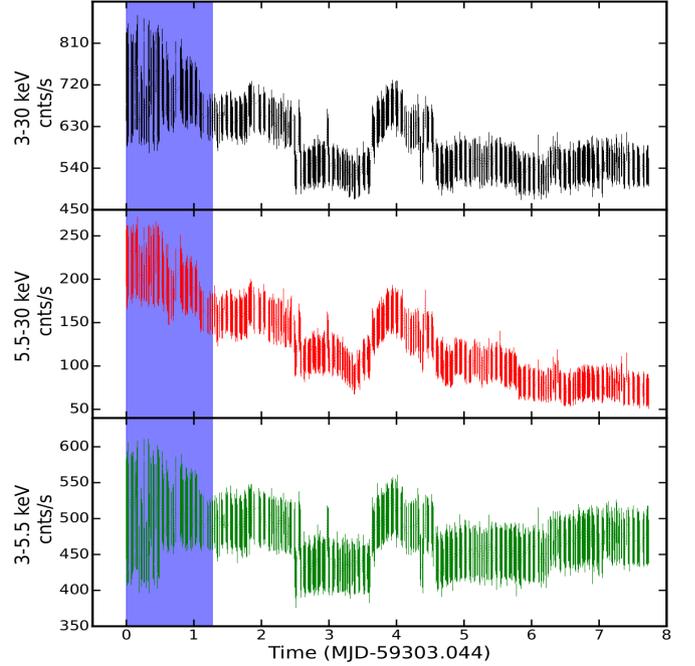}}};
\end{tikzpicture}
\caption {{\it AstroSat}/LAXPC20 light curves of GX\,339--4 in different energy bands for all 74 orbits, from the top, 3.0$-$30.0 keV (panel 1), 5.5$-$30.0 keV (panel 2), and 3.0$-$5.5 keV (panel 3), respectively. The bottom panel shows the Hardness ratio (5.5$-$30.0 keV / 3.0$-$5.5 keV). The blue shaded region shows the 11 orbits in which QPOs are observed. The `0' (zero) in the X-axis represents MJD=59303.044.}
\label{fig:LCurve}
\end{figure}

\autoref{fig:LCurve} shows the light curves of GX\,339--4 in different energy bands during the observation period. From the top, the panels in \autoref{fig:LCurve} correspond to energy bands 3.0$-$30.0 keV, 5.5$-$30.0 keV, and 3.0$-$5.5 keV, respectively. In the hard band (5.5$-$30.0 keV), flux has changed by a factor of 2, whereas in the soft band (3.0$-$5.5 keV), flux change is by a factor of 1.2. These changes indicate that the hard flux is more variable than the soft flux, which is quite expected, as the hard radiations are coming from the inner hot region, where temperature and the size of the corona can change significantly due to the change in accretion flow parameters \citep{ChakTitarchuk1995ApJ...455..623C} in a timescale of days. Alternatively, it has been observed that the change in rms-energy spectra in different spectral states can also be a probe to identify such flux changes, which may depend on the intrinsic variability of the corona \citep{BelloniEtal2011BASI...39..409B}. However, as we have short-term data, and it belongs to one spectral state only, verifying such a probe is beyond our scope. In addition, there is an alternating flux state \citep[AFS;][]{LiuEtal2022MNRAS.513.4308L} in all energy bands in the {\it Insight-HXMT} data for this source observed during 1-2 April 2021. The AFS looks similar to `Flip-Flop' profile (FFP) as observed for other sources \citep[][and references therein]{BogensbergerEtal2020A&A...641A.101B,JitheshEatal2019ApJ...887..101J} including GX\,339--4 \citep{Miy91}. This AFS (here timescale is $\lesssim$ 100 ks) might not be the same as the FFP in terms of their variability timescale. It is possible that some outflows/jet activities were triggered during that time, and the flux went down, as outflows/jets can carry some energy and mass from the inner corona region. However, some other effects, e.g., the change in mass accretion rate due to change in viscosity  \citep{Lyubarskii1997MNRAS.292..679L,MondalEtal2017ApJ...850...47M} or irradiation  \citep{KingRitter1998MNRAS.293L..42K,Mondal2020AdSpR..65..693M} can generate such fluctuations in the light curve. We note that either FFP, AFS, or X-ray variability in general in a short time interval can be explained due to the change in corona temperature and geometry, which are the effects of change in mass accretion rate \citep[][see for recent FFP study.]{MondalJithesh2023MNRAS.522.2065M}. In a follow-up work by Jana et al. (2023) is studying the spectra of all orbits to explain such X-ray variabilities of the source. 

\begin{figure}
\hspace{-0.5cm}
       \begin{tikzpicture}
\draw (0, 0) node[inner sep=0] {\raisebox{0.0cm}{\includegraphics[height=7.0truecm,trim={0.0cm 0.0cm 0.0cm 0.0cm}, clip]{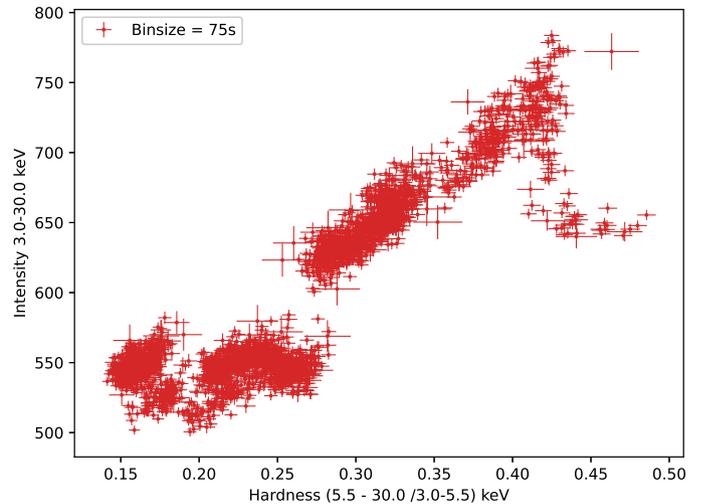}}};
\end{tikzpicture}
        \caption{Hardness intensity diagram is plotted during the {\it AstroSat} observation of the source. Based on the QPO properties including frequency, fractional rms, and Q-factor, the HID falls in the SIM branch of the full HID shown in \citet{peirano2023}.}
        \label{fig:hid}
\end{figure}

The HID during the {\it AstroSat} observation period is shown in \autoref{fig:hid}. Based on the QPO properties including frequency, fractional rms, and quality factor \citep[see also][]{MottaEtal2011MNRAS.418.2292M}, our estimated HID using {\it AstroSat} data falls in the soft intermediate (SIM) branch of the full HID, which has been shown by \citet{peirano2023} using {\it NICER} data. In the later sections, we confirm that the spectral state is SIM state (SIMS) using different QPO properties. We should mention that the authors used a different energy band to calculate the hardness ratio. Therefore, some discrepancies between the HIDs may arise. This is expected as hardness has no standard reference energy bands and changes with the availability of energy bands in different instruments.
  
\begin{figure}
\hspace{-0.2cm}
\begin{tikzpicture}
\draw (0, 0) node[inner sep=0] {\raisebox{0.0cm}{\includegraphics[height=7truecm,trim={0.0cm 0.0cm 2.0cm 1.8cm}, clip]{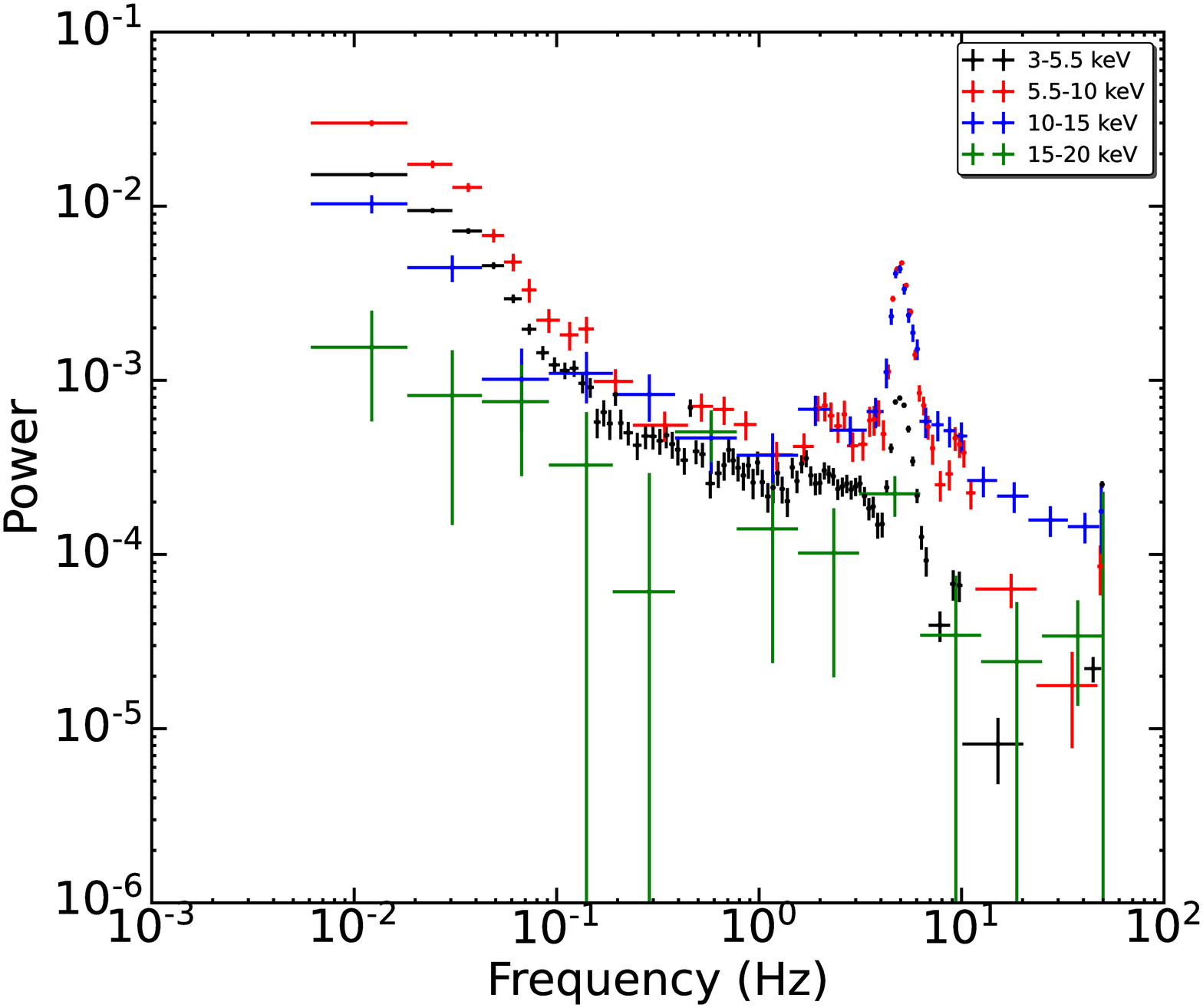}}};
\end{tikzpicture}
\caption{The power density spectra (PDS) of GX\,339--4 for different energy bands. All PDS are merged, combining all 74 orbits. Multiple Lorenzian components and a power-law component are used to fit the QPOs. The lower panel of \autoref{table:qpos} shows the QPO frequency estimated for the merged orbits for the energy band 5.5-10 keV and the Lorentzian model components. The power-law model is used to fit the broadband noise with normalization of $2.96\pm0.389 \times 10^{-4}$. A geometrical rebinning factor with step 0.02 is used for generating the PDS.}
\label{fig:qpoEDepend}
\end{figure}

Along with the flux variability, the light curve variabilities become prominent when we look at the PDS of the same light curves. \autoref{fig:qpoEDepend} shows the PDS in different energy bands. Interestingly, high energy bands ($>$ 20 keV) do not show any prominent QPOs, while in the low energy bands ($<$ 15 keV), sharp QPOs with harmonics and sub-harmonics are observed. The data above 20 keV are noisy, due to which QPOs were not detected in this energy range. However, it does not mean the QPOs are absent in the light curves beyond 20 keV. \citet{JinEtal2023ApJ...953...33J} observed QPOs above 50 keV for this source in {\it Insight-HXMT} data and other sources also showed the presence of QPOs at higher energy bands \citep[see][]{HomanEtal2001ApJS..132..377H,MaEtal2021NatAs...5...94M}. 
 The $\nu_{\rm qpo}$ observed from the combined 74 orbits in the energy band 5.5-10 keV are 2.14$\pm$0.08 ($\nu_{\rm qpo}^{\rm s}$), 5.06$\pm$0.01 ($\nu_{\rm qpo}^{\rm f}$), and 9.77$\pm$0.19 Hz ($\nu_{\rm qpo}^{\rm h}$). The other parameters of the QPOs fitting are shown in the bottom panel of \autoref{table:qpos}.  For the orbits 29756 and 29758, $\nu_{\rm qpo}^{\rm s}$ and $\nu_{\rm qpo}^{\rm f}$ follow a 1:2 ratio. However, the other two orbits do not. While we merged all orbit data, the QPO frequencies approximately satisfy the 1:2 ratio. Very recently, similar QPO frequencies were reported by \citet{peirano2023} for this source using {\it AstroSat} and {\it NICER} data. Authors used data of 7 orbits of {\it AstroSat} among 11 orbits that showed QPOs, and the inferred QPO properties and spectral type remained the same as we observed in this work.

\begin{table*}
%\small
\scriptsize
    \centering
    \caption{Best-fitted QPO frequencies and their properties. The first 11 rows show the QPO properties of all 11 orbits when the QPOs are observed. The last panel shows the QPO properties in the 5.5-10 keV energy band when all 74 orbits ($\sim 7$ days) are merged together.}\label{table:qpos}
    \begin{tabular}{{|c|c|c|c|c|c|c|c|c|c|c|c|c}}
    \hline
Orbit & Cnt Rate & Funda.  &   FWHM &    norm   &fractional&Q& Harmonic   &   FWHM &   norm & s-harmonic & FWHM & norm\\
(X=297) & (Cnts s$^{-1}$) & ($\nu_{\rm qpo}^{\rm f}$ Hz) & (Hz) &($10^{-3}$)  &rms (\%) & & ($\nu_{\rm qpo}^{\rm h}$ Hz) & (Hz) & ($10^{-4}$)&($\nu_{\rm qpo}^{\rm s}$ Hz)&(Hz)&($10^{-3}$)\\
\hline
X50 &$720\pm1$&5.32$\pm$0.02 & 0.72$\pm$0.06 & 42$\pm$2 &0.81$\pm$0.04&7.39$\pm0.62$& -& -& -&2.28$\pm$0.17 & 0.56$\pm$0.30 & 7$\pm$2 \\
X55 &$720\pm2$ &5.34$\pm$0.01 & 0.79$\pm$0.03 & 45$\pm$1 &0.88$\pm$0.02&6.76$\pm$0.26& -& -& -&2.77$\pm$0.05 & 0.75$\pm$0.13 & 11$\pm$1 \\
X56 &$685\pm2$ &5.32$\pm$0.02 & 1.00$\pm$0.05 & 28$\pm$1 &0.80$\pm$0.03&5.32$\pm$0.27& -& -& -&2.80$\pm$0.09 & 1.66$\pm$0.14 & 30$\pm$2 \\  
X58 &$722\pm3$ &5.57$\pm$0.02 & 1.02$\pm$0.06 & 40$\pm$2 &0.94$\pm$0.04&5.46$\pm$0.32& -& -& -&2.80$\pm$0.08 & 1.15$\pm$0.24 & 18$\pm$2 \\
X59 &$710\pm2$ &5.05$\pm$0.01 & 0.56$\pm$0.02 & 58$\pm$2 &0.85$\pm$0.02&9.02$\pm$0.32& 10.08$\pm$0.14 & 0.41$\pm$0.38 & 10$\pm$4 &-&-&-\\
X60 &$689\pm2$ &4.89$\pm$0.01 & 0.45$\pm$0.02 & 54$\pm$1 &0.74$\pm$0.02&10.87$\pm$0.48& - & - & - &-&-&-\\
X61 &$674\pm2$ &4.75$\pm$0.01 & 0.42$\pm$0.02 & 49$\pm$1 &0.69$\pm$0.02&11.31$\pm$0.54& 9.49$\pm$0.36 & 0.12$\pm$0.21 & 10$\pm$2 &-&-&-\\
X63 &$699\pm2$ &4.89$\pm$0.01 & 0.47$\pm$0.02 & 54$\pm$1 &0.76$\pm$0.02&10.40$\pm$0.44& 9.76$\pm$0.06 & 0.40$\pm$0.16 & 10$\pm$3 &-&-&-\\
X64 &$689\pm1$ &4.80$\pm$0.01 & 0.43$\pm$0.02 & 53$\pm$1 &0.72$\pm$0.02&11.16$\pm$0.52& 9.46$\pm$0.06 & 0.18$\pm$0.17 & 10$\pm$2 &-&-&-\\
X65 &$694\pm1$ &4.86$\pm$0.01 & 0.42$\pm$0.02 & 54$\pm$1 &0.72$\pm$0.02&11.57$\pm$0.55& 9.55$\pm$0.09 & 0.36$\pm$0.23 & 10$\pm$3 &-&-&-\\
X70 &$663\pm2$ &4.83$\pm$0.01 & 0.39$\pm$0.02 & 19$\pm$1 &0.42$\pm$0.02&12.38$\pm$0.64& - & - & -  &-&-&-\\ 
\hline
Merged &\multirow{2}{*}{$629\pm1$}&\multirow{2}{*}{5.06$\pm$0.01}&\multirow{2}{*}{0.91$\pm$0.03}&\multirow{2}{*}{70$\pm$2}&\multirow{2}{*}{1.26$\pm$0.03}&\multirow{2}{*}{5.56$\pm$0.18}&\multirow{2}{*}{9.77$\pm$0.19}&\multirow{2}{*}{2.50$\pm$0.64}&\multirow{2}{*}{10$\pm$3}&\multirow{2}{*}{2.14$\pm$0.08}&\multirow{2}{*}{0.69$\pm$0.26}&\multirow{2}{*}{50$\pm$10}\\
74 orbits&&&&&&&&&&&&\\
\hline
    \end{tabular}
    \label{tab:qpo_var}
\end{table*}

\begin{figure}
\hspace{-0.5cm}
       \begin{tikzpicture}
\draw (0, 0) node[inner sep=0] {\raisebox{0.0cm}{\includegraphics[height=8.5truecm,width=9.0truecm,trim={0.5cm 0.5cm 0.2cm 0.0cm}, clip]{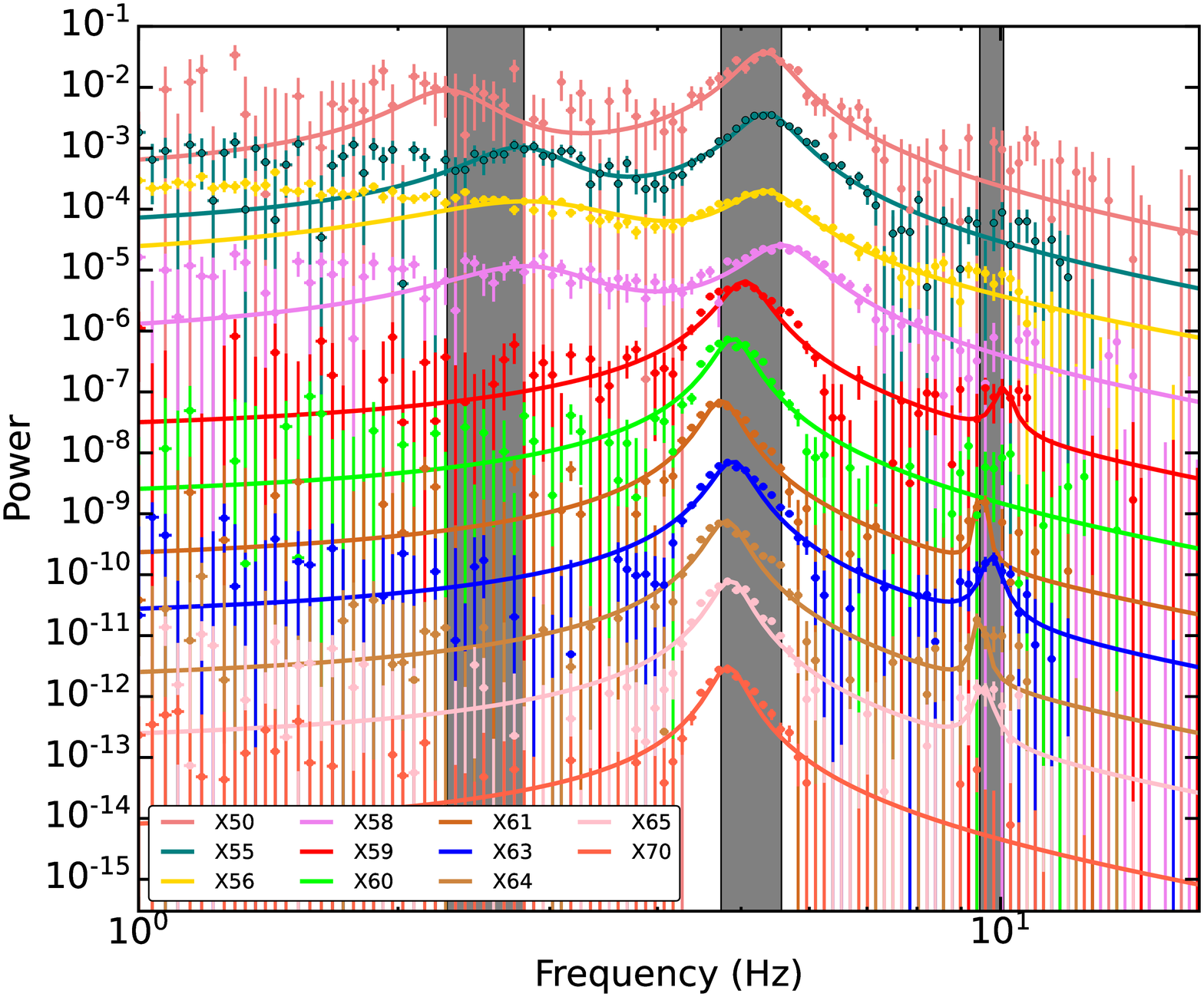}}};
\end{tikzpicture}
\caption{Orbit-wise variation of QPOs in the 3-25 keV energy band. `X' denotes the common prefix of orbit notation `297'. The power is scaled manually during plotting for visual clarity. The range of the sub-harmonic, fundamental, and harmonic peak frequencies are marked as \textcolor{gray}{grey} regions. The data values and errors are shown with '+' symbol, and the same color line represents the model.} \label{fig:qpoVarFit}
\end{figure}

\begin{figure}
\hspace{-0.5cm}
       \begin{tikzpicture}
\draw (0, 0) node[inner sep=0] {\raisebox{0.1cm}{\includegraphics[height=7.5truecm,trim={0.5cm 0.0cm 2.0cm 1.0cm}, clip]{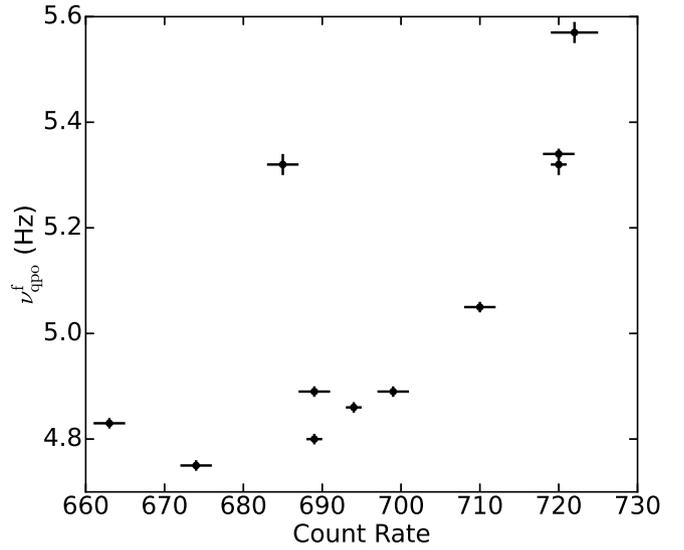}}};
\end{tikzpicture}
        \caption{Fundamental QPO frequency increases with increasing count rate in the 3-25 keV energy band during the observation period.}
        \label{fig:CntQpo}
\end{figure}

The orbit-wise (for 11 orbits) PDS fits are shown in \autoref{fig:qpoVarFit} and the corresponding QPO parameters are given in \autoref{table:qpos}. The data and their errors are presented with `+' symbol, and the same color line represents the fitted Lorentzian models. The vertical shaded regions (grey) show the shift in peak frequencies for all three QPOs ($\nu_{\rm qpo}^{\rm s}$, $\nu_{\rm qpo}^{\rm f}$, $\nu_{\rm qpo}^{\rm h}$), respectively. Moreover, we estimated the fractional rms and Q-values of the frequencies for all 11 orbits. Our estimated fractional rms is between 0.4\% and 0.9\% and the $\nu_{\rm qpo}^{\rm f}$ varies between 4.8-5.6 Hz. A series of studies showed that the type-B QPOs are limited to the range of 1–6 Hz, but the detections during high-flux intervals are concentrated in the narrow 4–6 Hz range. The centroid frequency appears positively correlated with source intensity rather than hardness \citep[see][]{CasellaEtal2005ApJ...629..403C,BelloniEtal2011BASI...39..409B,MottaEtal2011MNRAS.418.2292M}. If we compare our findings with the above classification, our data fall in the SIMS, and the QPO nature is type-B. Additionally, we have further checked the correlation between $\nu_{\rm qpo}^{\rm f}$ and intensity or count rate to compare with the above studies, which is shown in \autoref{fig:CntQpo}.

At the start of the observation, the sub-harmonic QPO appeared in a couple of orbits and then disappeared, but the harmonic was absent during that time. However, the opposite scenario was observed during the later time of the outburst. The appearance and disappearance of QPOs are intrinsic to the source, which can be due to the change in accretion flow parameters that are triggering such behavior, or it can be due to the launching of jets/mass outflows from the system \citep[see][and references therein]{FenderEtal2004MNRAS.355.1105F,SriramEtal2016ApJ...823...67S,MondalJithesh2023MNRAS.522.2065M}. The QPOs are only observed in the first 11 orbits (MJD 59303.06–59304.20) of the AstroSat observation, while the AFS is seen on MJD 59305–59306. Hence, there is no relation between AFS and the evolution of QPOs in this case, in comparison with the observation by \citet[][in Swift~J1658.2-4242]{BogensbergerEtal2020A&A...641A.101B}. 

The evolution of the QPO frequencies shows some interesting features, where either sub-harmonic or harmonic frequencies are present along with the fundamental. The evolution of QPOs and their harmonics can be explained using several popular models in the literature \citep[see][for a review]{IngramMotta2019NewAR..8501524I}. However, inferring the spectral state from QPO properties indicates that both spectral state and timing properties are interlinked. Therefore, the same model that explains the origin of QPOs should be able to fit the spectra. Such a physical connection between the spectral and timing properties is unlikely for the models discussed in the review. Therefore, it is suggestive and worth considering models which can explain both properties together. 
This motivates us to explain the evolution of the fundamental QPO frequency using the Propagating Oscillatory Shock model \citep[POS;][]{chakrabartiEtal2008A&A...489L..41C,NandiEtal2012A&A...542A..56N,Janaetal2016} including GX\,339--4 \citep[][and references therein]{DebnathElal2015MNRAS.447.1984D}, where an axisymmetric standing shock which is the boundary layer of the corona oscillates and produces such $\nu_{\rm qpo}^f$. The same shock that produces QPOs also decides spectral states during the outburst period of BHXRBs. In progressive days, as the source moves from a hard to soft state, the Keplerian disc moves inwards, and the cooling rate increases due to more soft photons getting scattered by the hot corona. Therefore, the corona shrinks, and QPO frequency increases. The electron number density and the temperature of the corona determine the cooling rate. However, explaining the harmonic or sub-harmonic frequencies requires some additional mechanisms, which can either be some perturbations to the axisymmetric shock and make it non-axisymmetric or the presence of more than one shock \citep{ChakrabartiAchaRyu2004A&A...421....1C}. As the source is observed in SIMS, it is clear that the dynamic corona or the centrifugal barrier has not completely disappeared, rather it moved inward due to an increase in mass accretion rate. In such a situation, one shock can break into fragments and oscillate in different frequencies \citep[see][for a review]{ChakrabartiEtal2009arXiv0903.1482C}. On the other hand, turbulence can produce a second shock along with the centrifugal barrier. As GX\,339-4 is a rapidly spinning black hole, and the turbulence has significant effects in originating LFQPOs \citep{Mondal2020MNRAS.492..804M}, it can be possible that two shocks oscillated in different frequencies and produced observed sub- or harmonic frequencies. However, performing numerical simulations in 3D can help in understanding such effects.

\begin{figure}
\begin{tikzpicture}
\draw (0, 0) node[inner sep=0] {\raisebox{0.0cm}{\includegraphics[height=7.5truecm,trim={0.0cm 0.0cm 1.5cm 1.2cm}, clip]{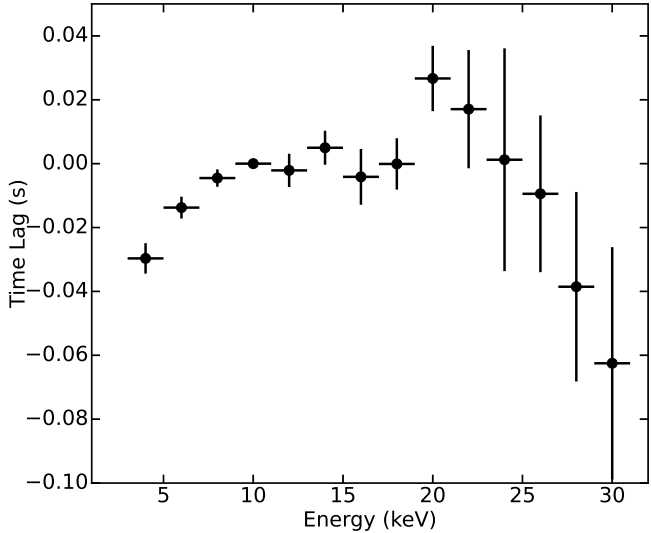}}};
\end{tikzpicture}
\caption{The energy dependent time lag estimated during the {\it AstroSat} observation period, considering $\nu_{\rm qpo}^f$ as the centroid frequency of the source. During estimating the lag at the frequency 5.06 Hz, we used a frequency range of 3-7 Hz.}
\label{fig:TLag}
\end{figure}

The QPOs and their evolution are directly linked with the size of the Comptonizing corona, whereas the time lag estimation not only gives the dynamic and thermodynamic properties of those regions but also about the energy-dependent physical processes that are responsible for the emission of photons from various regions of the disc \citep[see,][and references therein]{CuiEtal1997ApJ...484..383C,NowakEtal1999ApJ...517..355N,DuttaChakrabarti2016ApJ...828..101D}. In one of the frameworks of the Comptonization process, soft radiations get up-scattered by the hot electron cloud and reach the observer later than the soft photons, which are directly coming to the observer. In \autoref{fig:TLag}, we show the energy-dependent time lag spectrum of the type-B QPO of GX\,339-4. 
We find that the time lag is initially negative and increases with energy till $\sim$ 22 keV and then again negative. A similar lag behavior was observed for this source by \citet{JinEtal2023ApJ...953...33J}. The positive lag attributes to the Comptonization, while the negative lag may be explained by the jet or outflow \citep{Patra2019}. Such complex behavior of the lag spectrum may not be explained by a single mechanism such as viscosity fluctuation or the Comptonization process, it requires a detailed study using more broadband data. 

\begin{figure}
\hspace{-0.5cm}
\begin{tikzpicture}
\draw (0, 0) node[inner sep=0] {\raisebox{0.0cm}{\includegraphics[height=7.0truecm,trim={0.0cm 0.0cm 0.0cm 0.0cm}, clip]{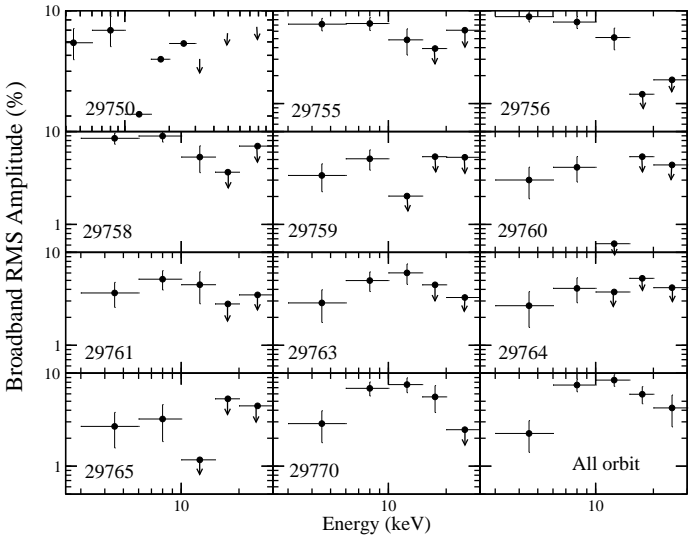}}};
\end{tikzpicture}
\caption{The rms spectra for all orbits when QPO is observed. Down arrow indicate the upper limit. The bottom right corner plot shows the rms amplitude for all 74 merged orbits. The rms amplitude values are estimated in the frequency range of 0.1-50 Hz.}
\label{fig:rms}
\end{figure}

\autoref{fig:rms} shows the rms spectra for each orbit when the QPO is observed. For each 11 orbit (that showed QPOs), we observed that the rms rises slowly, reaches maximum $\sim$10~keV, and then decreases. Therefore, it is reasonable to say that the photons having energy $\sim 10$~keV may contribute mostly to the QPOs \citep{vanderklis1989,vanderklis1994}, suggesting that the coronal emission is responsible for the QPOs. To further cross-check this finding, we have fitted the QPO in \autoref{fig:qpoEDepend} for different energy bands and noticed that Lorentzian normalization peaked for 5.5-10 keV band with values for all four bands are $(1.1, 6.9, 6.1, 4.0)\times10^{-3}$ respectively. This is in accord with other black hole X-ray binaries where rms peak is observed $\sim 10$~keV \citep[e.g.,][]{Alabarta2020,JanaEtal2022MNRAS.511.3922J}. Interestingly, in the case of a neutron star, the peak is observed at higher energy \citep[$>20$~keV, e.g., ][]{Wang2012}.

\section{Summary}

During the 2021 outburst of GX\,339-4, we found that the source showed significant variability in flux in a timescale of $\lesssim$ 100 ks. A detailed energy-dependent temporal study using {\it AstroSat} data shows that:

\begin{itemize}
    \item The source showed alternating flux profile (AFS) in both hard and soft energy bands. However, the hard energy band is found to be more variable than the soft band. The AFS can originate due to the jet/mass outflow activity or some local viscosity fluctuations in the disc.

    \item The soft energy band showed more prominent QPO along with its sub-harmonic and harmonic frequencies. In the hard energy bands, however, the QPO peaks become broader, and its components disappear. This can be due to a low S/N ratio in the data and may present in high-energy bands as well.

    \item Orbit-wise QPO study shows that the QPO frequency evolves with time. The sub-harmonics appear during the start of the observation and disappear at a later time, while the opposite scenario was observed for the harmonics.

  \item The evolution of QPO frequencies indicates that the size of the Comptonizing corona evolved with time, in agreement with the propagating oscillatory shock and multiple-shock scenario.

    \item The energy-dependent time lag study infers its association and origin from the Comptonizing corona. In our study, we found both positive and negative lag, whose origin can not be explained only using the Comptonization process. Some other complex processes might be responsible. To understand that, more broadband data is required to be analyzed. 

    \item In the rms spectra, the rms is observed to peak around $\sim$10~keV, suggesting that mostly photons with energy $\sim10$~keV are responsible for the QPOs.
\end{itemize}

\section*{Acknowledgements}
We thank the anonymous referee for making constructive comments and suggestions that improved the quality of the manuscript. We thank Prof. R. Misra of IUCAA for helping with time lag estimation. SM acknowledges Ramanujan Fellowship research grant (\#RJF/2020/000113) by SERB-DST, Govt. of India for this work. AJ and HK acknowledge the support of the grant from the Ministry of Science and Technology of Taiwan with the grand numbers MOST 110-2811-M-007-500 and MOST 111-2811-M-007-002. HK acknowledge the support of the grant from the Ministry of Science and Technology of Taiwan with the grand number MOST 110-2112-M-007-020 and MOST-111-2112-M-007-019.

%%%%%%%%%%%%%%%%%%%%%%%%%%%%%%%%%%%%%%%%%%%%%%%%%%
\section*{Data Availability}
The data used in this article are available in the ISRO's Science Data Archive for {\it AstroSat} Mission (\url{https://astrobrowse.issdc.gov.in/astro_archive/archive/Home.jsp}). The analysis techniques used in the paper can be shared on reasonable request to the corresponding author.



\bibliographystyle{mnras}
\bibliography{qpo-astrosat} % if your bibtex file is called example.bib
%\printbibliography
%\begin{thebibliography}{99}
%\end{thebibliography}

%%%%%%%%%%%%%%%%%%%%%%%%%%%%%%%%%%%%%%%%%%%%%%%%%%

%%%%%%%%%%%%%%%%% APPENDICES %%%%%%%%%%%%%%%%%%%%%

%\appendix
% \begin{figure}
% \hspace{-0.5cm}
%        \begin{tikzpicture}
% \draw (0, 0) node[inner sep=0] {\raisebox{0.0cm}{\includegraphics[height=8truecm,trim={0.5cm 0.5cm 0.2cm 0.0cm}, clip]{Figures/qpo_variation_range.pdf}}};
% \end{tikzpicture}
%         \caption{Orbit-wise variation of QPOs. `X' denotes the common prefix of orbit notation `297'. The power is scaled manually during plotting for visual clarity. The range of the sub-harmonic, fundamental, and harmonic peak frequencies are marked as \textcolor{red}{red}, \textcolor{gray}{grey}, and \textcolor{blue}{blue} regions, respectively.}
%         \label{fig:qpoVarRan}
% \end{figure}  


%%%%%%%%%%%%%%%%%%%%%%%%%%%%%%%%%%%%%%%%%%%%%%%%%%

% Don't change these lines
\bsp	% typesetting comment
\label{lastpage}
\end{document}